\documentclass[twocolumn,showpacs,prl,aps]{revtex4}
\usepackage{graphicx}
\begin{document}
\title{Generalised Shastry-Sutherland Models in three and higher dimensions}
\author{Naveen Surendran and R. Shankar}
\affiliation{The Institute of Mathematical Sciences, 
C.I.T. Campus, Chennai - 600113, India.}
\date{\today}

\begin{abstract}
We construct Heisenberg anti-ferromagnetic models in arbitrary dimensions that 
have  isotropic valence bond crystals (VBC) as their exact ground states. 
The $d=2$ model is the Shastry-Sutherland model. In the 3-d case we show that 
it is possible to have a lattice structure, analogous to that of 
$SrCu_2(BO_3)_2$, where the stronger bonds are associated with shorter bond 
lengths. A dimer mean field theory becomes exact at $d\rightarrow\infty$ and a 
systematic $1/d$ expansion can be developed about it. We study 
the Neel-VBC transition at large $d$ and find that the transition is 
first order in even but second order in odd dimensions.
\end{abstract} 

\pacs{75.10.Jm, 75.30.Kz, 75.50.Ee}

\maketitle

Recently there has been a renewed interest in the 2-d Shastry-Sutherland
model (SSM) \cite{shastry1}, owing to its physical realisation in
$SrCu_2(BO_3)_2$ \cite{kageyama,ueda}. The model has an exactly solvable
ground state. There exists some generalized antiferromagnetic Hamiltonians 
with exact ground states \cite{aklt,shastry2,caspers1,caspers2}.  

SSM was initially constructed
as a 2-d generalisation of the 1-d Majumdar-Ghosh model \cite{majumdar}.
Both models have a valence bond crystal (VBC) as the exact ground state.
Other such models, including a 3 dimensional one \cite{3d},
have since been constructed.

All the abovementioned models can be thought of as special cases of the
class of models which we will define below and refer to as generalised
Majumdar-Ghosh models (GMGM). Consider Hamiltonians of the form,
\begin{equation}
\label{gmgmham}
H = \sum_n J_n h_n
\end{equation}
Where the sum is over all possible triangles formed by the sites of the
lattice, $J_n$'s are arbitrary positive semidefinite couplings and $h_n$ are
given by,
\begin{equation}
\label{hamtriangle}
h_n ={ \bf S}({\bf r}_i).{ \bf S}({\bf r}_j) +
{ \bf S}({\bf r}_j).{ \bf S}( {\bf r}_k)
+{ \bf S}({\bf r}_k).{ \bf S}({\bf r}_i)
\end{equation}
${\bf S}({\bf r}_i)$, ${ \bf S}({\bf r}_j)$ and ${ \bf S}({\bf r}_k)$
are the spins at sites $ {\bf r}_i$, $ {\bf r}_j$ and $ {\bf r}_k$
respectively and $n$ labels the triangle formed by them. A particular 
${\bf r}_i$ can be a part of more than one $h_n$.

\begin{figure}
\includegraphics[width=.32\textwidth]{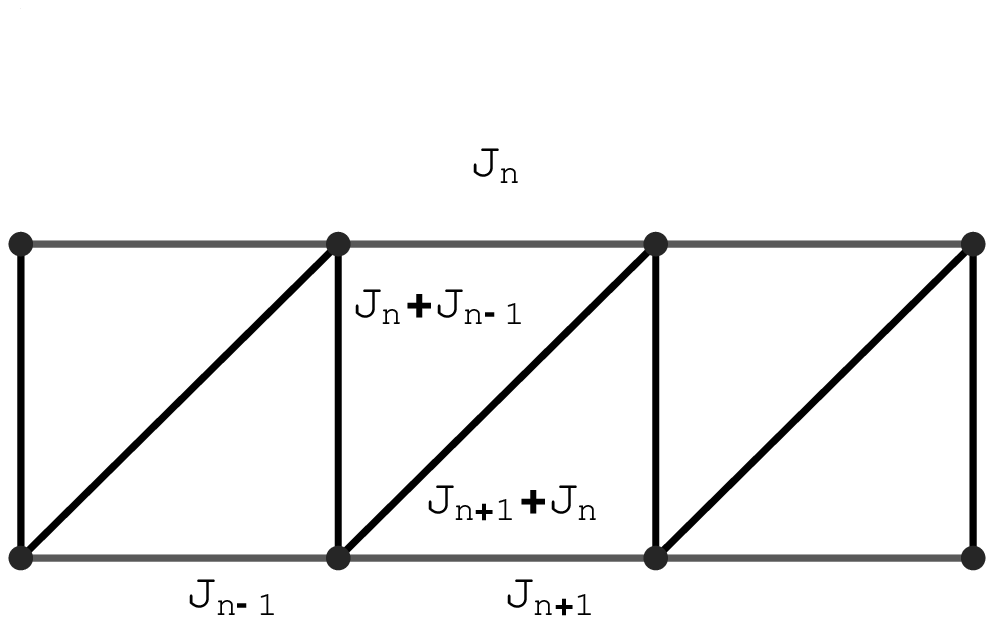}
\caption{\label{mgm} A generalised Majumdar-Ghosh Chain}
\end{figure}

It was noted in \cite{shastry1} that if, in the triangle corresponding
to $h_n$, two of the spins are forming a singlet, then the state will
be a ground state of $h_n$. It was also pointed out that $h_n$
could be more general than given above. It could be spin anisotropic, the
three terms could have different coefficients and the spin could be
arbitrary. In suitable parameter ranges the dimer
state will remain the ground state \cite{shastry1,kanter}. Thus, if it is
possible to cover the lattice with dimers in such a way that each of the 
triangles that appear in $H$ with a non-zero coupling contains a dimer 
then the state with singlets on all the dimers will be a simultaneous ground 
state of all $h_n$ and hence that of $H$.

\begin{figure}
\includegraphics[width=.32\textwidth]{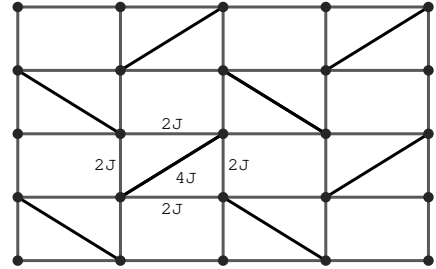}
\caption{\label{ssm} Shastry-Sutherland Model}
\end{figure}
The problem of constructing GMGMs then reduces to the purely geometric
one of assigning the non-zero couplings to the triangles such that a dimer
covering of the above type is possible. We will now give a class
of solutions to this problem which naturally generalises the SSM to arbitrary
dimensions. For simplicity, we will work with $h_n$ as in equation
(\ref{hamtriangle}) and with $s=1/2$.

\begin{figure}
\includegraphics[width=.32\textwidth]{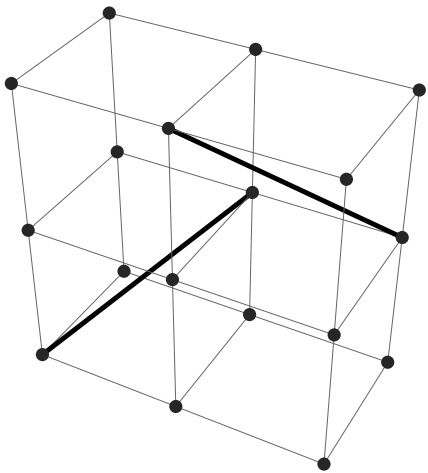}
\caption{\label{lattice} The 3-d lattice containing 18 sites, depicting the 
diagonals along which dimers are formed}
\end{figure}

We will first construct the 3-d model and then generalise to arbitrary
dimensions. We first set up the VBC and then build the Hamiltonian
around it. We take a simple cubic lattice and choose the dimers to lie along
the body diagonals. The body diagonals are assigned as follows. The sites
are denoted by, ${\bf x}= \sum_{\mu=1}^{3} x_\mu {\bf \hat{e}}_\mu$.
$x_\mu$'s take integer values and ${\bf \hat{e}}_\mu$ are three orthogonal
unit vectors. The spin ${\bf S}({\bf x})$ is paired to the spin
${\bf S}({\bf y}({\bf x}))$ where,
\begin{equation}
\label{diagonal}
{\bf y}({\bf x}) =  {\bf x} + {\bf D}({\bf x})
\end{equation}
\begin{equation} \label{diavec}
{\bf D}({\bf x}) =  \sum_{\mu=1}^{3} (-1)^{x_{\mu+1}} \hat{{\bf e}}_\mu
\end{equation}
where we define $ x_{3+1}\equiv x_1$. Two such body diagonals are shown in
FIG. \ref{lattice}.

Note that  ${\bf D}({\bf y})= - {\bf D}({\bf x})$ as it should be, since if
${\bf S}({\bf x})$ is paired with ${\bf S}({\bf y})$, then
${\bf S}({\bf y})$ should be paired with ${\bf S}({\bf x})$. All the
four body
diagonal directions occur in equal numbers and the VBC has cubic rotational
symmetry. It is not parity invariant, the other parity choice being given
by replacing $(-1)^{x_{\mu+1}}$ by $(-1)^{x_{\mu-1}}$ in equation (\ref{diavec}).

\begin{figure}
\includegraphics[width=.32\textwidth]{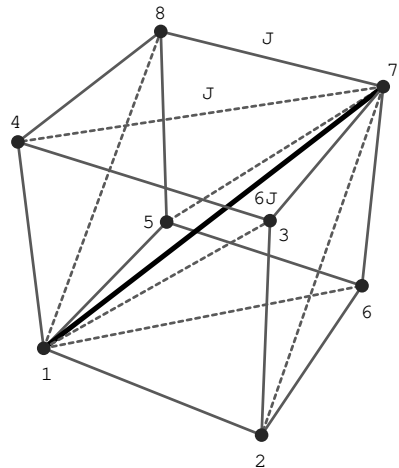}
\caption{\label{unitcell} A cube containing diagonal bond. All the bonds are 
shown}
\end{figure}

We now choose the triangles with nonzero couplings as follows. Equation
(\ref{diavec}) uniquely associates a body diagonal and hence a unit cube
with every dimer, $({\bf x},{\bf y})$. We give non-zero couplings to the
six triangles formed by these two sites and each of the other six sites that
belong to the cube as illustrated in FIG. \ref{unitcell}. Thus every such triangle has one edge, one face diagonal
and the body diagonal containing the dimer. This construction ensures that
every triangle has one dimer (along the body diagonal) and hence is the
exact ground state of Hamiltonians of the form given in equation
(\ref{gmgmham}).
The Hamiltonian can be explicitly written as,
\begin{equation}
\label{ham3d}
H=\sum_{{\bf x}} \sum_{\mu=1}^3 J({\bf x},\mu)h({\bf x},\mu)
\end{equation}
where,

\begin{eqnarray}
h({\bf x},\mu)&=& {\bf S}({\bf x}). {\bf S}({\bf z}({\bf x},\mu))
+{\bf S}({\bf z}({\bf x},\mu)).{\bf S}({\bf y})\nonumber\\
 & & {}+{\bf S}({\bf y}).{\bf S}({\bf x})\\
\label{edge}
{\bf z}({\bf x},\mu)&=&{\bf x}+ (-1)^{x_{\mu +1}}\hat{{\bf e}}_\mu
\end{eqnarray}
${\bf z}({\bf x},\mu)$ and ${\bf x}$ form the 3 edges emanating from ${\bf x}$ 
in the direction of the body diagonal ${\bf D}({\bf x})$ and {\bf y} is given 
by equation(\ref{diagonal}). $J({\bf x},\mu)$ is the coupling associated with
the triangle formed by ${\bf x},~{\bf y}({\bf x})$ and ${\bf z}({\bf x},\mu)$.

Consider the simplest case when all the couplings are equal. i.e.
 $J({\bf x},\mu)=J$. The triangle corresponding to $h({\bf x},\mu)$
contributes a strength $J$ to the edge it contains. Each edge is contained in 
exactly one triangle. Thus all edges have bond strengths $J$. Each triangle 
contributes a strength $J$ to the face diagonal that it contains. Half the 
face diagonals are contained in exactly one triangle each and the other 
half are not contained in any triangle. Thus half the face diagonals have bond
strength $J$ and the others have no bonds. FIG. \ref{unitcell} 
illustrates the situation. Finally, each triangle contributes $J$ to
the body diagonal. Half the body diagonals are contained in six triangles 
each and hence have bond strengths $6J$ and the other half have no bonds. 
See FIG.  \ref{lattice}. 

The generalisation to higher dimensions, $d=4,5,...$ is straightforward.
Simply replace 3 by $d$ in all formulas from equation (\ref{diagonal}) to
equation (\ref{edge}). All the $2^{d-1}$ body diagonal directions occur in equal numbers
in the $d$ dimensional VBC. The model is a simple hyper-cubic lattice with
bonds of strength $J$ along all the edges and along one of the face
diagonals of each $(d-1)$ dimensional face. There are also bonds of strength
$2dJ$ along one of the body diagonals of half the hyper-cubes. The construction
ensures that the VBC is the exact ground state of the model.

It can be seen that the model reduces to SSM at $d=2$, shown in FIG. \ref{ssm}.
The diagonals are  given by equation(\ref{diagonal}). The strength of the
bonds along the diagonals is $4J$. $d=2$ is a special case in that the
$(d-1)$ dimensional face diagonals are also the edges. Thus the strength along 
the edges is $2J$, ie. the strength of edge bonds are half that of the diagonal
ones. Thus we have recovered the 2-d SSM.

Now we come back to the $d=3$ case. As it stands, it is not very physically 
realistic since the stronger
bonds are between spins further apart. The same is true in the case of the 2-d
model. However, the structure of magnetic ions in $SrCu_2(BO_3)_2$ can be
obtained from the original theoretical lattice by moving the sites along the
diagonals that have bonds. The squares containing the diagonals deform to
rhombi and the body diagonal containing the dimer becomes shorter than the
edges. As we will see, the procedure generalises to 3-d and we can obtain an
analogous structure where the stronger bonds have shorter bond lengths.

\begin{figure}
\includegraphics[width=.32\textwidth]{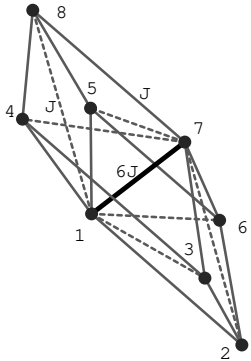}
\caption{\label{deformed} A deformed cube. Now the strongest bond is between 
nearest neighbour sites}
\end{figure}

We move all sites along the body diagonals with non-zero bond strengths. The
new sites are then,
\begin{equation}
\label{deform}
{\bf R}({\bf x})={\bf x} + \frac{s}{2}  {\bf D}({\bf x})
\end{equation}
where $s$ is a parameter and {\bf D}({\bf x}) is given by
equation(\ref{diavec}). The cube shown in FIG. \ref{unitcell} deforms to the 
rhombohedron shown in FIG. \ref{deformed}. Before the deformation the 
sites formed a simple cubic lattice. The Hamiltonian was however only 
symmetric under translations by two units. After deformation, the lattice 
periodicity is also halved. It remains a cubic lattice but with 8 sites in 
a unit cell.

\begin{figure}
\includegraphics[width=.32\textwidth]{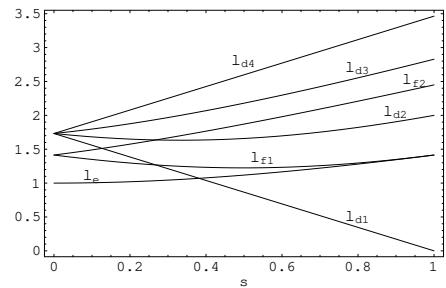}
\caption{\label{lengths} The various bond lengths as functions of
  $s$. $l_e$ is the length of the edges, $l_{f1}$ that of the face
  diagonals containing bond and $l_{f2}$ corresponds to face diagonals
  without bond. $l_{d1}$ is the length of the diagonals along which 
  dimers are formed and  $l_{d2}$,  $l_{d3}$ and  $l_{d4}$ correspond
  to body diagonals without bonds.}
\end{figure}

The lengths of the edges, face diagonals and body diagonals can be computed
and are plotted as functions of $s$ in FIG. \ref{lengths}. As we can see, for
$s$ more than around .7 or so, the edges and face diagonal with bonds become
almost equal in length, are longer than the body diagonal with the bond and
shorter than the other face diagonals and body diagonals. When $s$=1 the body 
diagonal becomes of zero length. The rhombohedron is then squashed into a 
hexagon lying in the plane orthogonal to the body diagonal.

We will now examine the models away from the exact ground state point
at $d\rightarrow \infty$. We put the bond strengths along the body diagonals
equal to $dJ_D/2$, along edges equal to $J_E/2$ and along the face diagonals
to $J_F/2$. Actually, the VBC is an exact eigenstate when $J_E=J_F\equiv 
J^\prime$ for all $J_D$ \cite{shastry1}. It is proven to be the ground state 
when $J^\prime \le 0.5J_D$. However, the ground state is Neel ordered for 
$J_F=0=J_D$. So there will be a phase transistion somewhere. The location 
and  nature of the transition at $d=2$ have been topics of much activity 
\cite{ueda,rrp,weihong1,mila,kawakami,weihong2}.

Since the VBC is the exact ground state in some parameter range for all $d$,
it is clear that mean field theory in terms of the spin variables fails even as
$d\rightarrow \infty$. This is because one of the bond strengths grows as
$d$ and so the interactions cannot be approximated by an average field.
However, if we take the dimers, i.e the two-spin systems on the body
diagonals to be the basic units, then each dimer interacts with $\sim d$ of 
the dimers around it with bonds of strength $\sim 1$. Thus the mean field 
theory in terms of the dimer variables is exact at $d\rightarrow \infty$ in 
this class of models. Perturbation theory around the mean field Hamiltonian 
then yields a systematic $1/d$ expansion for the fluctuations. We will now 
use this mean field theory to explore the physics at large $d$.

We label the spins as
${\bf S}_{I\alpha}$, where $I$ labels the positions of the centers of the
dimers and $\alpha = 1,2$ label the two spins that form the dimer. The 
Hamiltonian can be written as,
\begin{equation}
\label{dimerham}
H = \sum_I \frac{dJ_D}{4}({\bf S}_{I1}+{\bf S}_{I2})^2 +
\frac{1}{2}\sum_{I\alpha,J\beta}J_{I\alpha,J\beta}{\bf S}_{I\alpha}.
{\bf S}_{J\beta}
\end{equation}
Where $J_{I\alpha,J\beta}$ denote the edge and face diagonal bonds.
The mean field Hamiltonian is,
\begin{eqnarray}
\label{mfham}
H_{MF} & = & \sum_I \frac{dJ_D}{4}({\bf S}_{I1}+{\bf S}_{I2})^2 +
\sum_{I\alpha,J\beta}J_{I\alpha,J\beta}{\bf b}_{I\alpha}.{\bf S}_{J\beta}
\nonumber\\
&-& \frac{1}{2}\sum_{I\alpha,J\beta}J_{I\alpha,J\beta}{\bf b}_{I\alpha}.
{\bf b}_{J\beta}
\end{eqnarray}
The self consistency equations are then,
\begin{equation}
\label{sceq}
{\bf b}_{I\alpha} = \langle {\bf S}_{I\alpha} \rangle
\end{equation}

In different parameter regimes, the VBC as well as a variety of other
phases are possible. The $1/d$ expansion, which is valid in all the phases
can be used to analyse the phase diagram. In this paper we study the Neel-VBC 
transition at large $d$. There is a qualitative difference between odd and 
even dimensions and we treat them separately.

In even dimensions, the dimer lattice is bipartite and both the sites of a 
dimer have the same parity. The parity of a dimer is defined to be same as  
that of its sites. The Neel-state is described by the ansatz, 
${\bf b}_{I\alpha}=P_I b {\hat z}$, where $P_I$ is +1 on one sublattice and 
-1 on the other. Then the mean field Hamiltonian  is,
\begin{eqnarray}
\label{evenmf}
H_{MF}^{even} & = & \sum_I \Big( \frac{dJ_D}{4}({\bf S}_{I1}+{\bf S}_{I2})^2 
\nonumber\\
&+&P_I 2d J^\prime  b ( S_{I1}^{z} +  S_{I2}^{z}) 
+ 2d b^2 J^\prime \Big)
\end{eqnarray}
Where $J^\prime \equiv (J_E+J_F)/2$. For all values of 
$\alpha\equiv J^\prime/J_D$, all the dimers forming singlets, ie VBC, is a 
mean-field solution with $b=0$. The state has energy $0$. When $\alpha>1/2$, 
the Neel state is a solution with $b=1/2$ and has energy,
\begin{equation}
E^{Neel} = \frac{NdJ_D}{4}(1 - \alpha)
\end{equation}
Where N is the number of sites. However, this solution has lower energy than 
the VBC only when $\alpha > 1$. Thus we get a first order transition at 
$\alpha = 1$ at $d=\infty$. 

We have also computed the leading order correction to the ground 
state energy in the Neel phase by treating $(H - H_{MF})$ as a perturbation.
We get,
\begin{equation}
\label{neelen}
E = \frac{NJ_D}{4}\Big(d(1-\alpha)-\frac{\alpha}{4}\big(\frac{2\alpha}
{4\alpha-1} +\frac{1}{2}\big)\Big)
\end{equation}
At $d=2$, the transition now occurs at $\alpha \approx 0.8$. More 
sophisticated 
calculations at $d=2$ \cite{weihong1,ueda} put this number at 0.69. There are 
indications that the transition may be second order \cite{rrp} or that there 
may be an intermediate phase \cite{mila,kawakami,weihong2}. 

In odd dimensions the two sites of a dimer are not of the same parity. By 
convention, we assign $\alpha = 1$ for the odd site and $\alpha = 2$ for the
even site. Then the Neel ansatz is ${\bf b}_{I\alpha}=(-1)^\alpha b 
{\hat z}$. The mean field Hamiltonian is then given by,
\begin{eqnarray}
\label{oddmf}
H_{MF}^{odd} & = & \sum_I \Big( \frac{dJ_D}{4}({\bf S}_{I1}+{\bf S}_{I2})^2 
\nonumber\\
&+&2 d b \Delta J ( S_{I1}^{z} -  S_{I2}^{z}) 
+ 2d b^2 \Delta J \Big)
\end{eqnarray}
Where $\Delta J \equiv (J_E-J_F)/2$. In the ground state the dimer wave 
function is given by,
\begin{equation}
\label{wf}
\vert \Psi \rangle = cos\frac{\theta}{2}\vert 0,0 \rangle
+ sin\frac{\theta}{2}\vert 1,0 \rangle
\end{equation}
Where in $\vert l,m \rangle$, $l$ is the total spin and $m$ the $z$
component and $sin\theta=2\Delta J \vert{\bf b}\vert/(\sqrt{J_D^2+
(8\Delta J\vert{\bf b}\vert)^2})$. 
The self consistency equation (\ref{sceq}) reduces to,
\begin{equation}
\label{edsce}
\frac{4\Delta J}{\sqrt{J_D^2+(8\Delta J\vert {\bf b}\vert)^2}} = 1
\end{equation}
Thus the transition occurs at $J_D=4\Delta J$. The sub-lattice magnetization,
$\vert{\bf b}\vert$ goes continuously to zero at the transition as 
$\sim (1-(J_D/4\Delta J))^{1/2}$.
The interesting thing is that unlike in even
dimensions, the transition point depends on the difference of $J_E$ and $J_F$. 
So the VBC can occur at relatively low values of $J_D$. 

The different physics in the even and odd dimensions arises from the fact 
that the two spins on a dimer belong to the same
sublattice in former and on opposite ones in the latter.
Consequently, the dimer wavefunction in the 
VBC and in the Neel state have the same value of $S^z (=0)$ in odd dimensions 
whereas in
even dimension $S^z=0$ in the VBC but $S^z=\pm 1$ (on odd and even
sublattices) in the Neel state. Since the $S^z$ symmetry is unbroken in both
the phases, the mean field Hamiltonian always conserves it. Thus in even
dimensions the VBC state cannot smoothly transit to the Neel state and we
get a first order transition whereas in odd dimensions it can and we get a
second order transition. 

A remark about the scaling of the diagonal bond is in order here. For the 
dimer mean-field theory to be valid in both the phases, the diagonal bond 
has to scale as $d$, so that $d$ scales out of $H_{MF}$ of equations 
(\ref{evenmf}, \ref{oddmf}). But numerically, the critical value of the
diagonal bond could be small. For example, in odd dimensions the critical 
value of diagonal bond is proportional to 
$\Delta J \equiv (\frac{J_E - J_F}{2})$ and can be made arbitrarily small 
by suitable choice of $J_E$ and $J_F$.

The mean field equations can be solved in the presence of an external magnetic
field ${\bf B}$. The mean-field Hamiltonian in the presence of 
${\bf B}=B{\hat {\bf z}}$ is given by, 
\begin{eqnarray}
\label{bmfham}
H_{MF} & = & \sum_I \frac{dJ_D}{4}({\bf S}_{I1}+{\bf S}_{I2})^2 +
\sum_{I\alpha,J\beta}J_{I\alpha,J\beta}{\bf b}_{I\alpha}.{\bf S}_{J\beta}
\nonumber\\
&-& B\sum_{I\alpha}S_{I\alpha}^z-\frac{1}{2}\sum_{I\alpha,J\beta}J_{I\alpha,J\beta}{\bf b}_{I\alpha}.
{\bf b}_{J\beta}
\end{eqnarray}

We concentrate on the regime where the VBC is the ground state,  in the 
absence of ${\bf B}$. A state in which one of the dimers is excited to a 
triplet with $S^z=1$ and the rest are all in singlets
is a self-consistent solution of the mean-field Hamiltonian of 
equation(\ref{bmfham}), with ${\bf b}_{I\alpha}=0$ for those dimers in
singlets and ${\bf b}_{I\alpha}= \frac{1}{2}{\hat z}$ for the one in triplet. 
The energy of this state is $dJ_D/2-B$.
Thus as we increase the strength of the magnetic field, when $B>dJ_D/2$,
it is energetically favourable to excite as many dimers as possible into 
triplets pointing along ${\bf B}$, but no two of them being connected 
by a bond. 

In even dimensions, using the fact that the lattice formed by the dimers is 
bipartite, we shall now show that the maximum fraction of dimers that can be 
excited without any two of them being connected by a bond is $1/2$. 

Let us start with isolated dimers with no bonds connecting them. Now we 
add edge bonds such that every dimer is connected to one and only one other
dimer. This can be done as follows. As mentioned before, 
both the sites of a dimer have the same parity. Each coordinate of the two 
sites will be differing by $+1$ or $-1$. Choose an even dimer and pick out 
the site with odd $x_1$-coordinate. Put the edge bond from this site in the
positive $x_1$-direction. This way every even dimer can be uniquely connected 
to one and only one odd dimer. Suppose these were the only bonds present. Then 
the configuration that maximizes the number of triplets without two of them 
being connected is where in every pair of connected dimers one is put in a 
singlet and the other in a triplet. Now the original model can be obtained by
adding the remaining bonds. But in the presence of additional bonds also, each
of the pairs that were initially connected can at the most have one triplet. 
Thus the maximum fraction of dimers that can be excited without any two of 
them being connected by a bond is $1/2$. This upper bound can be satisfied by 
putting triplets on one of the sub-lattices and singlets on the other. This
state is a mean-field solution with ${\bf b}_{I\alpha}=0$ for those dimers in
singlets and ${\bf b}_{I\alpha}= \frac{1}{2}{\hat z}$ for those in triplets.  

The state with the remaining singlets also being excited into triplets is
also a mean-field solution and has energy $\frac{N}{2}(dJ_D/2+dJ^\prime-B)$ 
above the half-magnetized state. Thus when $B>dJ_D/2+dJ^\prime$ all the 
remaining dimers are excited into triplets and the system becomes fully 
magnetized. 

When $d$ is odd, the dimer lattice is not bipartite. We still assign parity
to dimers as follows. In every dimer one site will be odd and the other even.
The parity of the dimer is defined as the parity of the $x_1$-coordinate of 
the even site. It can be seen that out of the $4d$ dimers that are connected
to a particular dimer only $4$ are of the same parity. Since the critical $B$,
at which it is energetically favourable to have isolated singlets, scales as 
$d$, this difference between odd and even dimensions is insignificant at 
large $d$. 

For both the cases, there is a plateau at 1/2 magnetization for 
$dJ_D/2<B<dJ_D/2+dJ^\prime$. At stronger fields, the system is fully 
magnetized. The 1/2 plateau corresponds to all the
dimers on even sites in the triplet and the ones on the odd sites in the
singlet state. At $d=\infty$, the dimers have only nearest neighbour 
interactions. Previous work \cite{ueda,rrp,momoi} indicates that the other 
fractions are due to longer range interactions induced by fluctuations. Thus 
it is reasonable that only the 1/2 plateau occurs at $d=\infty$.

To summarise, we have constructed $d$-dimensional models, $d\ge 2$, that are
natural generalisations of the Shastry-Sutherland model. These are
Heisenberg antiferromagnets on hypercubic lattices with bonds of strength
$dJ_D/2$ along half the body diagonals, $J_E/2$ along all the edges and 
$J_F/2$ along half 
the $(d-1)$-dimensional face diagonals. They have isotropic VBCs as their exact
ground states at $J_D \ge 2J_E=2J_F$. A dimer mean field theory is exact in
the $d\rightarrow \infty$ limit and a $1/d$ expansion can be developed about
it. In this limit, the Neel-VBC transition is first order in even dimensions 
and occurs at $\alpha = 1$. When leading order ($1/d$) corrections are 
included, the transition shifts to $\alpha=0.8$ at $d=2$. In odd dimensions 
the transition is second order and occurs at $J_D=4\Delta J^\prime$. In 
presence of an external magnetic field, ${\bf B}$, in both even and odd 
dimensions, the system has a  1/2 plateau from $B=dJ_D/2$ to 
$dJ_D/2+dJ^\prime$ and becomes fully magnetised at higher fields.

At $d=3$, we have shown that it is possible to have a crystal structure
where the stronger bonds have shorter bond lengths. Moreover, the mean field
theory indicates that the transition to the VBC phase occurs at relatively 
small values of $J_D$. Thus a physical realisation of such a system seems 
feasible.

\end{document}